# Mode Coupling and Breathing Oscillation in Partially Magnetized Cross-Field Plasmas

Jong Yoon Park[1] and June Young Kim[2†]

[1]*Department of Nuclear Engineering, Seoul National University, Seoul 08826, Korea*
[2]*Department of AI Semiconductor Engineering, Korea University, Sejong 30019, Korea*

[†] Corresponding author
E-mail: juneyoungkim@korea.ac.kr

We report on investigations of mode coupling between rotating spokes during the onset of the breathing oscillation. Demonstrating the existence of nonlinear coupling between the sporadic spokes and the breathing oscillation, we suggest the oscillating azimuthal electric field as the energy source for additional ionization within the plasma. Our results indicate that intermittent three-wave coupling is a possible mechanism for triggering low-frequency breathing oscillations in partially magnetized cross-field plasma.

Instabilities and subsequent plasma state evolution are universally observed phenomena in plasma physics, spanning vast astrophysical realms to meticulously controlled laboratory settings [1−4]. They are especially important in fields such as fusion energy [5,6], space propulsion [7,8], and semiconductor processing [9,10], where sudden transitions in plasma states present formidable challenges. Managing these challenges involves controlling excessive heat flux in fusion reactor walls, mitigating orbital deviations during space missions, and safeguarding against wafer damage in semiconductor manufacturing. Understanding these dynamic instabilities has become increasingly vital owing to the advancement of technical innovation. The complex mechanisms underlying plasma generation lead to the emergence of various free energy sources, including plasma density and temperature gradients [11,12], velocity shear [13,14], and beam components [15,16]. This complexity emphasizes the need to gain a comprehensive understanding of nonlinear coupling phenomena among different instabilities. Consequently, theoretical [17−20] and experimental [21−24] investigations in this area have expanded to explore the captivating realm of nonlinear wave coupling phenomena.

While the understanding of instability physics and its extension to nonlinear phenomena has predominantly focused on fully magnetized plasmas, a growing interest exists in exploring these phenomena in partially magnetized plasmas, as applied in space thrusters and semiconductor processes [25−29, 38−41]. For partially magnetized plasmas in crossed fields, the instabilities deviate from those observed in large linear devices or nuclear fusion contexts [25−34]. These plasmas exhibit distinct attributes, including magnetization and temperature disparities between electrons and ions, and electric fields substantially influenced by boundary conditions. Notably, in long-wavelength-dominated scenarios, the instability is characterized by the rotation of a region (i.e., a rotating spoke) with enhanced light emission, moving at velocities below the E×B velocity and displaying a frequency range of 30–60 kHz [34−37]. Additionally, the presence of a partially ionized state with a low degree of ionization enables gas dynamics to directly contribute to the phenomenon of breathing oscillations, observed as discharge current fluctuations in the few kHz range. These oscillations are crucial for determining device stability [42]. Various theories have been proposed to explain these phenomena [43−48], often discussing them in the context of intense ionization at the source or linking them to plasma energy and transport [43].

Contemporary theoretical models underscore the intricate nonlinear behavior inherent in these instabilities. Key findings within these models spotlight phenomena such as intensified electron heating [38], amplified ionization via electron vortex formation [39], turbulence energy cascades [40], and the progression from a linear gradient-drift-driven instability to its nonlinear ionization wave counterpart [41]. While the theoretical interpretations concerning the nonlinear behavior of the instabilities are inconsistent, experimental evidence regarding the precise nature of these mechanisms is limited. In particular, rotating spokes and breathing oscillations have traditionally been treated as independent research topics. This distinction has overlooked the potential nonlinear coupling between these two instabilities, posing a significant limitation to understanding the mechanisms of instabilities in partially magnetized plasmas. We argue that this lack of theoretical agreement stems from a deficiency in experimental evidence detailing the nonlinear characteristics and coupling between instabilities in partially magnetized plasmas.

This letter demonstrates for the first time the three-wave coupling between sporadic spokes and breathing oscillation in partially magnetized cross-field plasmas and their causal relationship. Leveraging advanced experimental capabilities, our study created an optimal environment for observing three-wave coupling using a cylindrical hot cathode Penning source to identify eigenmodes with distinct azimuthal mode numbers, notably free from MHz range interferences [49,50]. We also measured radially propagating breathing oscillations [51]. Using these well-measured data, the distinct patterns between interacting local azimuthal modes and the emergence of global breathing are experimentally confirmed. The partially magnetized cross-field plasma source utilized in our study is illustrated in Figure 1. The Penning apparatus comprises two main sections: a relatively hot plasma column (spanning $0 - 8$ mm from the radial center) and an extraction region. The plasma column defines a cylindrical area between two axial boundaries: a negatively biased hot cathode with an outer shield (right side) and an anti-cathode (left side). The cylindrical source wall is grounded, establishing the reference potential for the entire plasma source system [52].

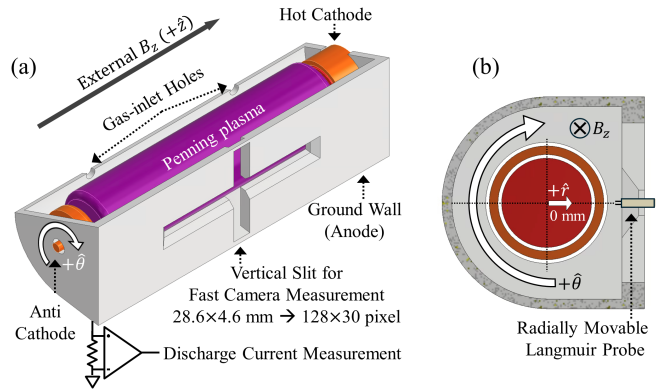

FIG. 1. Cross-section of the partially magnetized source as observed from (a) an angled top view and (b) the side with a radially movable Langmuir probe. The vertical slit in (a) is used for fast camera measurements. The radial center of the hot cathode and anti-cathode is set to the center. Each arrow in (b) indicates the axial ($+\hat{z}$), radial ($+\hat{r}$), and azimuthal ($+\hat{\theta}$) directions.

We utilized a two-tip probe with a tip spacing $d_{tip}$ of 1.4 mm (each tip has a radius of 0.15 mm and a length of 1.4 mm). The floating voltage $V_f$ signal from one tip of the two-tip probe was used to obtain the radial profile of the time-averaged frequency spectra and measure the azimuthal mode number [Fig. 2]. Floating voltage was detected using a radially movable Langmuir probe connected to an oscilloscope via a 1 MΩ resistor. The sporadic dynamics of breathing oscillations and rotating spokes were verified using a fast camera and simultaneous measurement of discharge current $I_{dis}$, without installing a Langmuir probe [Fig. 3].

The fast camera (Phantom V711) was installed to capture visible light changes in the plasma with a spatial resolution of −18 to 18 mm via a vertical slit with 128 × 34 pixels. It operated with a temporal resolution of 4.76 μs (210 kframe/s) and a shutter time of 4.3 μs, providing sufficient resolution to cover the instability dynamics. We conducted measurements facing the slit to avoid significant losses of magnetic field-aligned electron. The ion saturation current $I_{isat}$ was measured using one tip of the two-tip probe, while $I_{dis}$ was simultaneously measured to analyze their spectrograms [Fig. 4]. A probe voltage of −80 V was applied using a battery to ensure proper collection of $I_{isat}$. $I_{dis}$ oscillations were measured with a differential probe through a 0.5 Ω non-inductive resistor connected between the ground and anode of the discharge circuit [Fig. 1]. Measurements over 200 ms captured non-stationary behaviors in frequency spectrograms. The setup detected up to 11,764 oscillation cycles, with a maximum observable period of approximately 17 μs.

Initially, we aimed to understand the fundamental characteristics of the instabilities. Experiments were conducted to observe the oscillation of $V_f$. Throughout the experiments, the conditions were maintained at an average $I_{dis}$ of 1.7 A, a discharge voltage of 45 V, and an argon pressure of 66 mPa in the source and axial magnetic field, $B_z$, of 128 G. Frequency spectra were obtained using fast Fourier transforms (FFTs) on the complete signals from a floating probe over a duration of 200 ms. In Fig. 2(a), the radial profile of the FFT results of $V_f$ is presented. Three distinct frequency peaks were observed in the frequency spectra. Dominant fluctuation frequencies of approximately less than 10 kHz, 30 kHz, and 60 kHz remain nearly consistent across the radial direction, emphasizing the characteristic eigenmode of the system [Fig. 2(b)]. Figure 2(c) illustrates the radial variation of the fluctuation levels for each mode, with the magnitude of the fluctuation peaking at around 9 mm, and all modes displaying a similar profile. Using the Beall technique [53] on $V_f$ with a two-tip Langmuir probe, azimuthal mode numbers corresponding to frequencies below 10 kHz, 30 kHz, and 60 kHz were identified as 0, 2, and 3, respectively [small figure in Fig. 2(a)]. These frequency components of 30 and 60 kHz have been identified as rotating spokes driven by collisionless Simon-Hoh instability [52]. In this experiment, we denote rotating spokes with a large amplitude at approximately 30 kHz as the large spoke (LS) and rotating spokes with a small amplitude at approximately 60 kHz as the small spoke (SS). Our recent studies have demonstrated the radial propagation characteristics of low-frequency oscillations (less than 10 kHz) [51]. Accordingly, these low-frequency oscillations can be identified as breathing oscillations (BO), a phenomenon commonly observed in other cross-field sources [42−48].

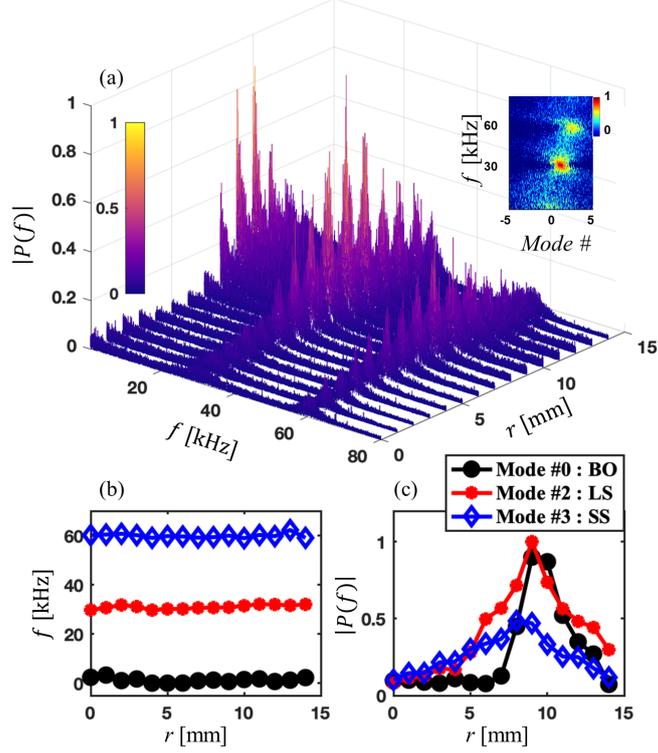

FIG. 2. The radial profile of (a) the frequency spectra for the $V_f$ from the Langmuir probe at $B_z$ of 128 G; (b) $f$ of each mode; and (c) $|P(f)|$ of each mode. $r$ and $|P(f)|$ denote the radial distance from the center of the source and the single-sided amplitude spectrum of frequency $f$, respectively. Mode # indicates the azimuthal mode number of each frequency component. The small figure in (a) displays the Beall plot at 14 mm. The $x$-axis represents the azimuthal mode number, and the color bar indicates the normalized power spectral density.

We analyzed the measurement results from the fast camera at $B_z$ of 159 G to gain further insights into the internal structure of the plasma over time. By eliminating the mean intensity of each pixel, we processed the images to represent fluctuating light emission. The original fluctuating image was processed into three separate images. The original fluctuating images without digital filtering are located on the left side (marked as 1) in Figs. 3(a) and (b). From the digital band pass filtering of the original images, further images showing internal modes were obtained; column 2 for BO to capture the fluctuations with frequencies less than 20 kHz, column 3 for LS (30−40 kHz), and column 4 for SS (60−70 kHz). Note that azimuthally rotating spokes and breathing oscillations are clearly seen in the fast camera measurements, and their overall characteristics are consistent with the electrical diagnostics. We defined the sudden decrease in brightness at frequencies below 20 kHz as the onset of *breathing*, and the period immediately preceding this event as the *pre-breathing* phase, as shown in column 2 of Figs. 3(a) and (b).

We select one of the pixels in images $\tilde{\gamma}_{55}$ indicated by a small black dot in column 1 of Fig. 3(a) to clarify its temporal behavior. The spectrogram of $\tilde{\gamma}_{55}$ is shown at Fig. 3(c); the *pre-breathing* and *breathing* phases are outlined by blue and magenta dashed boxes, respectively. From this spectrogram, we can observe the sporadic nature of each mode. The *breathing* phase coincides with the sudden decrease in $I_{dis}$, as shown in Fig. 3(d). Contrary to the widely recognized characteristics of spokes, they exhibited a sporadic pattern of formation and dissipation. The sporadic dynamics of modes indicate the absence of a specific, predictable cycle. This sporadic nature has led to the hypothesis that nonlinear wave coupling is actively occurring within the source.

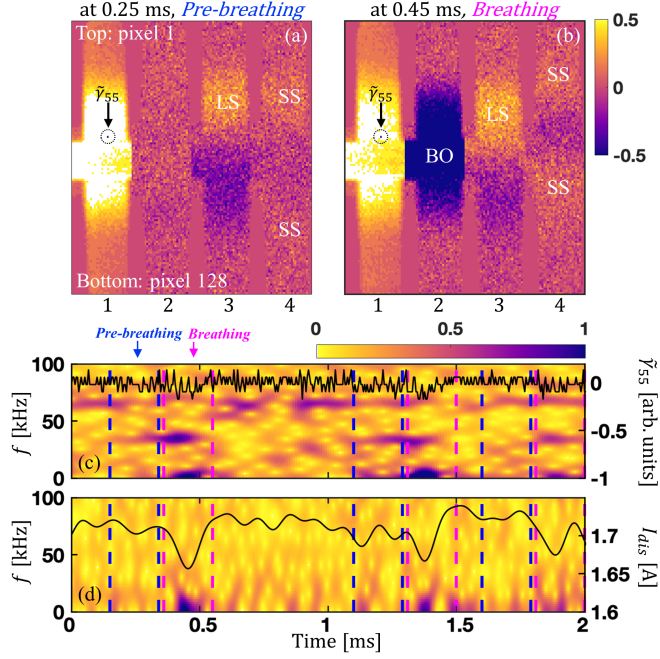

FIG. 3. Fast camera measurement results. Four filtered images at the (a) *pre-breathing* and (b) *breathing* phases. (c) The spectrograms of $\tilde{\gamma}_{55}$ and (d) $I_{dis}$.

A high order spectral analysis was performed to validate the hypothesis made in Fig. 3. The degree of nonlinear coupling among three different waves was assessed via bicoherence analysis of the oscillating azimuthal electric field $\tilde{E}_{field}$ achieved by the two-tip probe ($E_{field} = (V_{f,1} - V_{f,2})/d_{tip}$), where $V_{f,1}$ and $V_{f,2}$ indicate the two signals measured by the floating two-tip probe. In the experiment, the mean $E_{field}$ in the *pre-breathing* phase was measured as 460 V/m and 440 V/m in the *breathing* phase. For the analysis of $\tilde{E}_{field}$, these mean values of $E_{field}$ were removed. This analysis explores the phase relationship between two arbitrary waves ($f_1$ and $f_2$) and their corresponding summed frequency wave ($f_1 + f_2$). The results of a bicoherence analysis range from 0 to 1, where 0 indicates no coupling between frequencies and 1 signifies strong coupling [54,55]. This analysis involved ensemble averaging across numerous statistically analogous instances, totaling 359 cases for the *pre-breathing* and *breathing* phases, effectively eliminating arbitrary or coincidental phase matching. The formula for calculating the bicoherence $b_w^2(f_1, f_2)$ can be found elsewhere [54,55]. In the *pre-breathing* phase (before the breathing event), nonlinear interactions were predominantly observed among spokes (e.g., $f_1 \sim 60$ kHz and $f_2 \sim 30$ kHz) [Fig. 4(a)]. However, during the onset of the breathing oscillation, a distinct nonlinear interaction was evidently manifested between the spokes ($f_1 \sim 30$ kHz or 60 kHz) and the breathing oscillation ($f_2 \sim$ below 10 kHz), as captured by the auto bicoherence analysis [Fig. 4(b)]. The significance of this coupling phenomenon was further validated by the summed wavelet bicoherence $\sum b_w^2(f_1, f_2)$ [54,55]. This result demonstrates the nonlinear interaction of the breathing oscillation below 10 kHz and the small and large spokes. Such a coupling pattern exceeds the statistical significance level, affirming the physical validity of the nonlinear coupling [Fig. 4(c)]. The notable new findings from this analysis include the emergence of a new wave at 90−100 kHz and its amplification during the *breathing* phase, indicating the mode coupling and nonlinear effects, as the frequency of the new wave matches the sum of the frequencies of the existing two modes (30 kHz + 60 kHz = 90 kHz).

Moreover, having confirmed the nonlinear coupling through bi-coherency analysis, we were prompted to closely analyze the nonlinear power transfer across different frequencies, a phenomenon frequently observed during mode coupling. This can be calculated using the Ritz method [56]. The formula for calculating the total nonlinear spectral power transfer $\sum T_f$ is detailed in [56,57]. For this study, frequency is utilized as a substitute for wavelength. Taylor's hypothesis is verifiably satisfied (i.e., the lifetime $\tau_L$ of the $\tilde{E}_{field}$ must exceed the transit time $\tau_P$ of the $\tilde{E}_{field}$ between the two points), as $\tau_L$ and $\tau_P$ are approximately 2.0 $\mu s$ and 0.4 µs, respectively. The result is shown in Fig. 4(d). The $\sum T_f$ increase in regions below 10 kHz and 90−100 kHz and its decrease in the 30 and 60 kHz regions are notable. This indicates that the $\tilde{E}_{field}$ energy in LS and SS is released while the $\tilde{E}_{field}$ energy in frequencies below 10 kHz and in the new wave at 100 kHz is absorbed. Noting that the outflow direction of $\tilde{E}_{field}$ energy in the spokes does not mean that the energy is completely depleted; rather, it is continuously produced and consumed during the discharge. Given the sporadic nature of various modes in the source, nonlinear power transfers occur frequently. This result raises a critical question about the role of the released $\tilde{E}_{field}$ energy in the source. As a substantial $\tilde{E}_{field}$ energy outflow is observed, we anticipate that the released $\tilde{E}_{field}$ energy acts as free energy, contributing to changes in internal parameters.

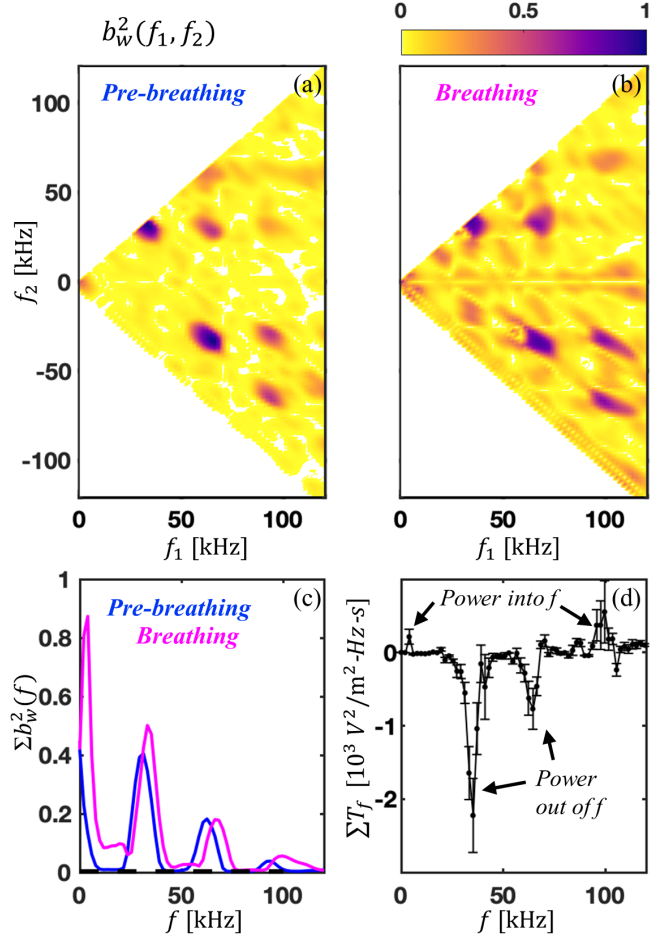

FIG. 4. Wavelet bicoherence results in the (a) *pre-breathing* and (b) *breathing* phases. $b_w^2(f_1, f_2)$ in arbitrary units represents squared wavelet bicoherence. (c) Summed wavelet bicoherence $\Sigma b_w^2(f = f_1 + f_2)$ of the above results. The black dashed line in (c) indicates the significance level of the analysis. (d) Total nonlinear spectral power transfer $\sum T_f$ in arbitrary units as a function of frequency during the dissipation of small spokes and the onset of breathing oscillation. The data from six radial positions ($r = 9$ to $14$ mm with $1$ mm interval) were averaged. The regions where power flows into and out of a specific frequency $f$ are indicated by black arrows.

The sporadic light emission of the internal mode, as shown in Figs. 3(a) and (b), along with the intermittently released $\tilde{E}_{field}$ energy, prompted us to estimate internal parameters, including temporal changes in neutral gas and plasma density, as well as local $\tilde{E}_{field}$. One of the novel characteristics of the Penning source is that $I_{dis}$ is proportional to the neutral pressure, allowing it to be used as a pressure gauge [58,59]. In our Penning source, the linear relation between neutral pressure $P_{source}$ and $I_{dis}$ is guaranteed under our experimental conditions ($P_{source} = 166 I_{dis} - 215$). To acquire statistically significant data of $P_{source}$, total of 57 breathing cases were obtained and ensemble-averaged. The time evolution of $P_{source}$ was then tracked, as shown in Fig. 5(a). In the breathing phase, a sharp decrease in neutral pressure is expected, which corresponds to a sudden drop in light intensity, as shown in Fig. 3(b). By using ensembled $I_{isat}$ from the same breathing cases and the relation $I_{isat} = 0.61 n_i A_{tip} u_B$, the time-varying density of electrons $n_e(t)$ was obtained (where $n_i$ is the ion density, $A_{tip}$ is the probe tip area, $u_B$ is the Bohm velocity, and assuming quasi-neutrality ($n_i = n_e$) and an electron temperature of 5 eV). Subsequently, the density of each mode (namely, $n_{e,BO}$, $n_{e,LS}$ and $n_{e,SS}$) was derived by digitally filtering $n_e(t)$, as shown in Figs. 5(b−d). Specifically, $n_{e,BO}$ is band-filtered below 20 kHz, while $n_{e,LS}$ and $n_{e,SS}$ are band-filtered within the 30−40 kHz and 60−70 kHz ranges, respectively.

One noticeable feature is that during the *breathing* phase, an increase in $n_{e,LS}$ is observed, eventually leading to an overall increase in $n_e$. Assuming that changes in the neutral pressure $\Delta P_{source}$ during the *breathing* phase (12.9 mPa, indicated by an arrow in Fig. 5(a)) are caused by electron-neutral impact ionization, the corresponding increase in the electron density $n_{e,add}$ is calculated as $4.06 \times 10^{17}$ m$^{-3}$ using the simple relation $\Delta P_{source} = n_{e,add} k_B T_{gas}$. The gas temperature $T_{gas}$ is assumed to be 0.2 eV, estimated by ion temperature measurement results in a similar set-up and conditions [60]. The changes in electron density $\Delta n_e$ during the *breathing* phase are obtained by calculating the difference between the initial density $n_{e,initial}$ and the maximum density $n_{e,max}$, as shown in Fig. 5(b). Notably, $\Delta n_e$ is measured as $2.65 \times 10^{17}$ m$^{-3}$, which is comparable to the expected change in $n_{e,add}$. This indicates that the ionization process is actively ongoing during the *breathing* phase while instantaneously consuming neutral gas. While the additional ionization could be contributed to LS as well as other components, precise partitioning among them was not confirmed in this study. Then, as $P_{source}$ recovers to its next *pre-breathing* phase, $n_e$ and $n_{e,LS}$ decreased. These tendencies of increased ionization are only observed during the *breathing* phase.

We analyzed the time evolution of $\tilde{E}_{field}$, used for nonlinear coupling analysis. Data from 74 breathing events were ensemble averaged, and the $\tilde{E}_{BO}$(filtered below 20 kHz), $\tilde{E}_{LS}$(band-filtered within the 30−40 kHz) and $\tilde{E}_{SS}$(band-filtered within the 60−70 kHz) are depicted in Figs. 5(e−g). Contrary to the increase in $n_e$ observed during the phase transition from *pre-breathing* to *breathing*, shown in Figs. 5(b−d), there is a slight decrease in local $\tilde{E}_{field}$, as illustrated in Fig. 5(e). Notably, reductions in the intensities of both $\tilde{E}_{LS}$ and $\tilde{E}_{SS}$ are measured during the phase transition, which correspond to the trends of nonlinear power transfer in the LS and SS frequency bins shown in Fig. 4(d). The observed time lag between the azimuthal electric field and plasma density variations has not been observed in other theoretical and experimental studies. After the *breathing* phase, the recovery of neutral pressure to its original levels and the subsequent decrease in density indicate that this ionization was driven by a transient supply of energy, as no additional external power was supplied during the Penning discharge. A drop in the $\tilde{E}_{field}$ after the *breathing* phase is also observed. We expect that this phase marks the beginning of the next wave-related events, though this is beyond the scope our current study.

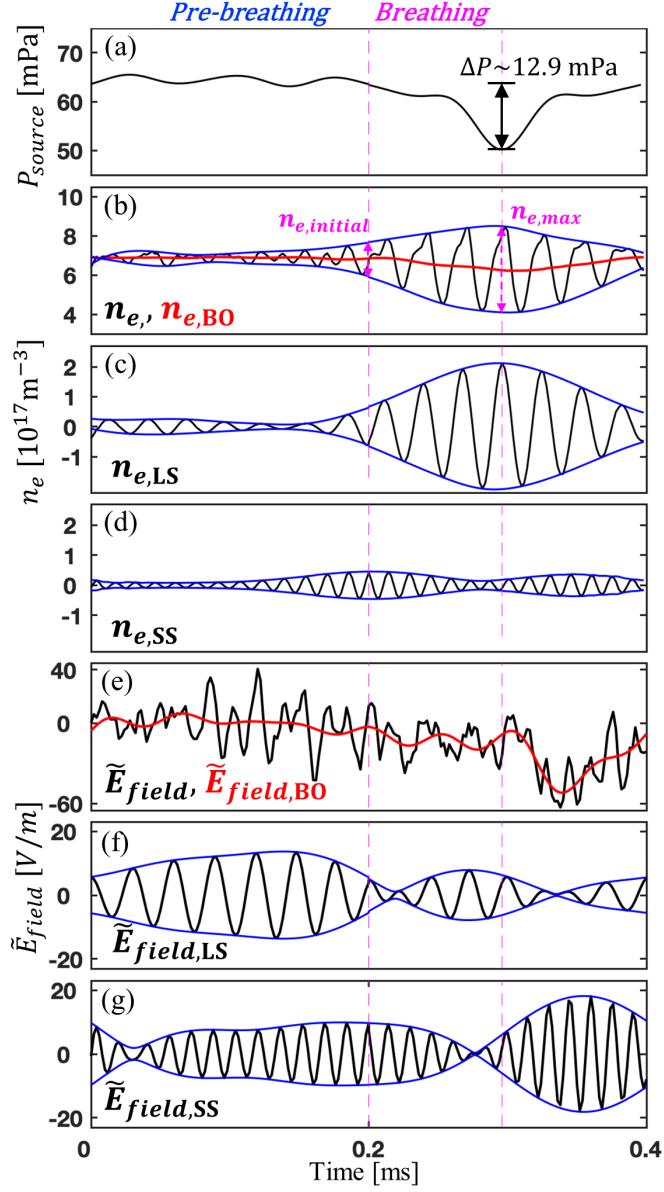

FIG. 5. Time-varying evolution of (a) $P_{source}$, (b) $n_e$ (black) and $n_{e,BO}$ (red), (c) $n_{e,LS}$, (d) $n_{e,SS}$, (e) $\tilde{E}_{field}$ and $\tilde{E}_{field,BO}$, (f) $\tilde{E}_{field,LS}$ and (g) $\tilde{E}_{field,SS}$. The blue lines in (b−d, f, g) indicate the amplitude envelope for a given waveform. The probe is located at $r$ = 10 mm. All data is ensemble averaged, with a total 57 breathing cases for (a−d) and 74 breathing cases for (b−d).

Eventually, we suggest a possible energy source facilitating the breathing oscillation mechanism by highlighting the intermittent strengthening of azimuthal $\tilde{E}_{field}$ in the spokes led to the breathing oscillation. Given the investigated nonlinear mode coupling and resultant changes in plasma parameters, we emphasize that a significant outflow of $\tilde{E}_{field}$ power from the spokes' bins during the *breathing* phase does not simply vanish; rather, it can be transformed into free energy. This free energy is not constrained to maintaining the oscillations of specific electric field modes and thus can be redirected to other processes, redistributing energy within the plasma, which in turn can enhance ionization rates. In addition, during the *breathing* phase, the super-positioned electric field oscillation $\tilde{E}_{coupled}$ during coupling can accelerate the plasmas. The combined electric fields from both modes (LS and SS) can induce instantaneous acceleration of electrons leading to ionization. This is likely because the real-time evolution of the electric field is highly spiky for both modes (data not shown here as all are ensemble averaged). Additionally, results in Fig. 5($f-$ g) show high values of $\tilde{E}_{field,\text{LS}}$ and $\tilde{E}_{field,\text{SS}}$ during the *pre-breathing* phase. Moreover, the intensities of $\tilde{E}_{field,\text{LS}}$ and $\tilde{E}_{field,\text{SS}}$ increase reciprocally during the *breathing* phase, indicating dynamic changes in the local electric field.

In conclusion, the nonlinear coupling observations collectively presented during intermittent breathing modes motivate phenomenological hypotheses for its trigger mechanism as follows: (1) two sporadic spokes are created in the Penning source; (2) the spokes undergo mode coupling as they are in harmonic relation, resulting in nonlinear coupling and energy transfer; (3) as the field energy in the spokes is transferred to other domain, it causes additional ionization; and (4) in the lab frame, the reduced light emission caused by the instantaneously decreased neutral pressure appears as '*breathing*'. Our study not only provides pioneering insights into the nonlinear interplay between coherent spokes and breathing oscillations in partially magnetized cross-field plasmas but also enhances our understanding of inherent instabilities in cross-field sources, thereby offering clear directives for future research and technological applications. Moreover, as mode coupling and the resulting phenomena observed in this study are actively researched in various plasma conditions including high-temperature plasmas [61], they enable detailed and accessible research on nonlinear coupling phenomena in small devices, which are challenging to study in larger devices such as Tokamaks. Additionally, more detailed experiments and research are required to address further questions, such as why only the $n_{e,\text{LS}}$ increases during the *breathing* phase and what the threshold conditions are for the *breathing* phase in a given mode coupling.

**Acknowledgments**

This research was supported by the National Research Foundation of Korea (NRF) grant funded by the Korean Government (MSIT) (RS-2023-00208968, NRF-2019R1A2C2089457, NRF-2021M3F7A1084418).